# LECAR'S VISUAL COMPARISON METHOD TO ASSESS

# THE RANDOMNESS OF BODE'S LAW: AN ANSWER

Vladimir Pletser

*Institut d'Astronomie et de Géophysique G. Lemaitre, Université de Louvain, 1348 Louvain-la-Neuve, Belgium.*
*Present address: Technology and Engineering Centre for Space Utilization, Chinese Academy of Sciences, PITC-325-1, No 9 Dengzhuang South Road, Haidian District, Beijing 100094, China*
*E-mail: Vladimir.Pletser@csu.ac.cn*

**Abstract**
The usual main objection against any attempt in finding a physical cause for the planet distance distribution is based on the assumption that similar distance distribution could be obtained by sequences of random numbers. This assumption was stated by Lecar in an old paper (1973). We show here how this assumption is incorrect and how his visual comparison method is inappropriate.

## 1. Introduction
The usual main objection against any attempt in finding a physical or dynamical explanation of the planet distance distribution is based on the assumption that similar distance distributions could be obtained by sequences of random numbers. Lecar stated this assumption, based on a limited set of results of Dole's computer simulations (Dole, 1970), in a short letter to Nature (Lecar, 1973).
Dole (1970) generated planetary-like systems by injecting small nuclei into a gas and dust nebula. The semi-major axis $a$ and the eccentricity of the initial nuclei injection orbit were chosen at random. Nuclei would grow into proto-planets by accreting dust and gas, if their mass $m$ and temperature were respectively high and low enough. Proto-planets would coalescence in case of crossing orbits or if coming within a critical interacting distance $d = a\sqrt[4]{\frac{m}{1+m}}$. Some of Dole's obtained planetary systems were similar to the solar planetary system in terms of spacing of orbits and size of individual planets.
From this, Lecar argued that the spacing ratio expressed in Bode's planetary distance law could be generated by sequences of random numbers subject to the constraint that adjacent planets cannot be "too close to each other". The physical reason of the constraint was that, if two planets were too close to each other during the accretion process, they would coalesce or cease to grow because of competition for the same material, as previously discussed by Dermott (1972).
Further, Lecar showed in his Figure 1 logarithmic plots of distances of seven Dole's computer-generated systems against increasing integers, next to the actual planetary system, which the reader was invited to recognize. By simple visual comparison, it was difficult to differentiate the planetary system from Dole's systems. From this, Lecar concluded "… *that this offers an equally satisfactory rationalization of Bode's mnemonic.*"

If the physical reason for the "closeness not too close" condition is perfectly correct and obvious, Lecar's visual comparison method is limited, his conclusion based on Dole's results is incorrect and his arguments are inconsistent.

## 2. Comparing the incomparable
*2.1 The visual comparison method*
One is baffled by the over-simplistic and un-quantitative proposed "pick and go" approach in Lecar's paper. It can be summarized as follows: "Try to pick up the good result and if you cannot, it means that the good result is similar to the other results". So simple: no need for figures, numbers, parameters, characteristics about these systems, you just have to guess visually the answer.

We suggest the reader to try this and attempt to pick out the planetary system plot among the four plots of Figure 1.

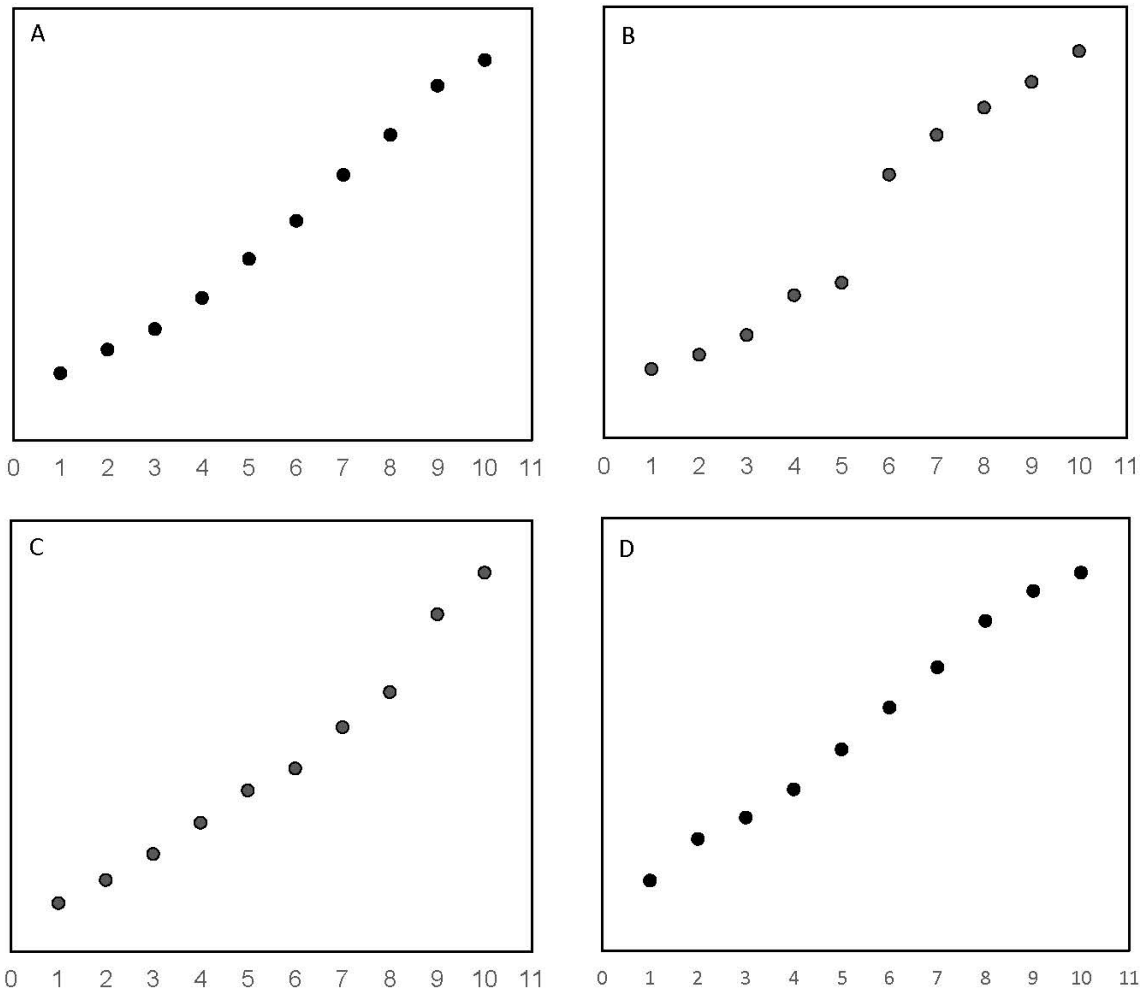

Figure 1: Plots against increasing integers of: (1) the logarithm of the planets semi-major axis, (2) the USA population every ten years between 1850 and 1940, (3) the sorted logarithm of $^{241}Am$ concentrations in surface intertidal sediments in ten sites near Sellafield, North of England (Mackenzie et al. 1987), and (4) the sorted telephone numbers of town halls of ten districts of Brussels, Belgium.

By simple visual comparison, the plot A seems to have the best accuracy, followed respectively by plots D, C and B. Is there any link between these distributions? Can we say from this comparison that the planetary distance distribution is similarly represented by the evolution of the US population in the last centuries, or by sorted logarithm values of nuclear pollution in the North of England (Mackenzie et al., 1987), or by sorted phone numbers of town halls in Brussels?

*2.2 Careful choice of Dole's random systems*
Dole (1970) conducted 200 simulation runs resulting in computer generated planetary-like systems. Twenty out of the 200 planetary-like systems were displayed in four figures in schematic plots of planet distances and sizes. Among these twenty systems, two systems had eight planets, five had nine, twelve had ten, and one had eleven.
In his paper, Lecar showed in Figure 1 logarithmic plots of distances of seven systems carefully chosen among Dole's twenty systems, similar to the actual planetary system and with the most regular distance distribution (see Figure 2), i.e. with ten planets[1] and an average spacing ratio close to the one of the planetary system.

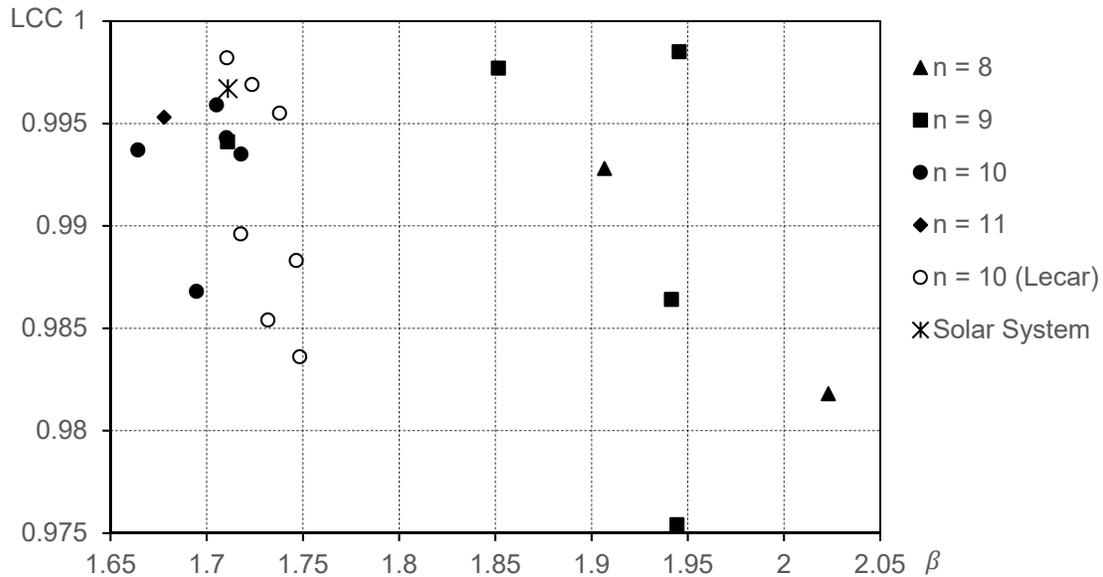

Figure 2: Distribution of the averaged spacing ratios $\bar{\beta}$ between planets and the Linear Correlation Coefficient (LCC) values of linearized exponential regressions versus increasing integers for Dole's displayed 20 systems having 8 (triangle), 9 (square), 10 (circle) and 11 (diamond) planets (open circles for the seven systems of 10 planets chosen by Lecar). Values of $\beta$ and LCC for the actual Solar System are indicated by an asterisk.

---

[1] Although Pluto was moved from the family of planets to the group of planetesimals in 2006 by the IAU, Lecar's arguments were made in the seventies when the planetary system counted nine planets and the asteroid belt.

Figure 1 of Lecar's paper displayed these seven systems next to the actual planetary system, which the reader was invited to recognize. A simple visual comparison was obviously not sufficient to differentiate between the eight displayed plots of ten dots in approximate straight lines, mainly because all eight systems had Linear Correlation Coefficients (LCC) close to or greater than 0.99. One can ask why any of the other 13 systems showed by Dole were not chosen or why the remaining 193 systems were not considered. Is it because they were not regular enough in terms of more or less constant spacing ratios to be exhibited next to the plot of the Solar System? Wouldn't it have been better to consider all 200 Dole's systems and either choose at random seven or more among these 200 systems or to make an accurate statistic on these 200 planetary-like systems and then make a comparison with the actual Solar System?

*2.3 Linear fit*
It is obvious that a more or less accurate linear fit will always be found through ten numbers generated at random and sorted in ascending order, in function of increasing integers. It will be easily found, after several trials, that most of the LCC's range between 0.9 and 0.999. Can we deduce from this that the present planetary spacing ratio can be similarly expressed by this simple process? Furthermore, imposing in the random generation process the constraint that two successive numbers cannot be "too close to each other" will increase artificially the fit accuracy, as the allowed region of existence of a next random value is reduced.

*2.4 More quantitative answers*
A more complete statistical analysis has been performed (Pletser, 2017), based on comparison of ratios of consecutive distances obtained randomly by three different generators (uniform, normal and exponential) and subjected to the "closeness not too close" condition with actual planetary masses in Dole's critical distances $d$. Results show that the average spacing ratios and the LCC of linearized exponential regressions of distances versus increasing integers are smaller than the observed actual values of the solar planetary system.
A similar analysis has been done (Pletser, 1987) with protoplanets higher mass at the end of the accretion phase, i.e. increased up to the Solar abundance, instead of present planets mass in Dole's critical distances $d$. It yielded similar results.
Furthermore, random systems with masses of the Jovian, Saturnian and Uranian satellite were also investigated (Pletser, 1988). Again, similar results were found. i.e. random systems having on average mean values of their spacing ratios and LCC's smaller than the actual systems.

## 3. Conclusion
We therefore conclude that the distance relation of the present planetary system, particularly the mean spacing ratio, cannot be expressed similarly by sequences of random numbers subject to the constraint of "closeness not too close" with the planets mass. This condition is certainly necessary but not sufficient to explain the exponential spacing observed in the present planetary system.
Furthermore, the visual comparison method used by Lecar and the arbitrary selection of computer-generated systems are definitely not appropriate. In Figure 1, panel A corresponds to caption (2) (with a LCC = 0.9945), panel B to caption (4) (LCC = 0.9818), panel C to caption (3) (LCC = 0.9852), panel D to caption (1) (LCC = 0.9967). "*Comparaison n'est pas raison[2]*".

---

[2] This French proverb can be translated as "Comparison is not reason".